\documentclass[final,5p,times,twocolumn,nofootinbib]{elsarticle}
\biboptions{numbers,sort&compress}

\usepackage{graphicx}
\usepackage{slashed}
\usepackage{tikz,mathpazo}
\usetikzlibrary{shapes.geometric, arrows}
\usepackage[none]{hyphenat}
\usepackage[english]{babel}

\usepackage{amssymb}
\usepackage{lipsum}
\usepackage{amsmath}

\journal{Physics Letters B}

\begin{document}

\begin{frontmatter}



\title{Inhomogeneous SU(2) gluon matter under rotation}


\author[first]{Yin Jiang}
\affiliation[first]{organization={Physics Department, Beihang University}, addressline={37 Xueyuan Rd}, city={Beijing}, postcode={100191}, country={China}}

\begin{abstract}

In this work a rotating SU(2) gluon system have been studied with the dyon ensemble in dilute limit. By solving the rotation-modified Yang-Mills equation we have obtained rotational corrections to the so-called dyon solutions with arbitrary centers to $\mathcal{O}(\omega^2)$ order and the corresponding semi-classical potential. The radial position dependent deconfinement temperature have been obtained by minimizing the semi-classical potential in both real and imaginary angular velocity cases. Although without the $\omega$-dependent coupling constant the critical temperature behaves different from the lattice simulation at each radial position as the rotation goes faster, its radial dependence is qualitatively the same as the lattice. That is in the real velocity case the outer layer will deconfine more difficult while the reverse is true in the imaginary velocity case.

\end{abstract}




\end{frontmatter}




\section{Introduction}
\label{introduction}
Rotation-induced phenomena have been extensively examined since measurements~\cite{STAR:2017ckg} of hyperon polarization at the final stage of Quark Gluon Plasma (QGP) evolution, which is generated through relativistic heavy ion collisions. Both global and local polarization measurements on hadrons~\cite{STAR:2018gyt, Xia:2018tes, Liang:2019pst, STAR:2020xbm, STAR:2021beb,  Ryu:2021lnx, Deng:2021miw, Li:2021zwq} have spurred numerous investigations into fermionic systems under rotation, typically associated with chiral restoration and anomalous transport~\cite{Huang:2017pqe, Abramchuk:2018jhd,  Wei:2018zfb, Wu:2019eyi, Ayala:2019iin, Ivanov:2020wak, Deng:2020ygd, Yi:2021ryh}. The rotating gluon system and deconfinement transition have recently garnered increased attention due to the gluon's larger spin for polarization and the dominant population of soft gluons in QGP. To explore a pure gluon system, lattice simulation of quantum chromodynamics(QCD) is the primary choice. However, in the rotating system, the pure gluon action becomes complex, thereby introducing a novel sign problem. By employing an imaginary angular velocity, lattice simulations can work and have yielded intriguing results~\cite{Chernodub:2020qah, Chernodub:2022wsw, Chernodub:2022veq, Braguta:2021ucr, Braguta:2021jgn, Braguta:2023iyx, Yang:2023vsw}.

The sign problem arising from rotation in lattice QCD renders the use of effective models in theoretical approaches more significant. In the context of a static, pure SU(N) gluon system, non-perturbative characteristics such as the well-known color confinement are typically addressed through semi-classical approaches, given the prevailing belief that the non-trivial topology of the principal bundle $e^{i A(x)}$ is paramount in the underlying theory. This conceptual understanding has prompted pivotal research into various semi-classical solitons of the non-Abelian Yang-Mills equation. To formulate an effective model, these solitons are often employed as the fundamental degrees of freedom, with their associated interactions derived by integrating out quantum fluctuations about them~\cite{Kraan:1998sn, Kraan:1998pm, Lee:1998bb, Diakonov:2005qa, Lawrence:2023woz, Escobar-Ruiz:2017uhx, Diakonov:2002qw, Diakonov:2004jn, Diakonov:2007nv, Schafer:1996wv, Kharzeev:2007jp}. Consequently, selecting an appropriate model for the static system and subsequently adapting it to incorporate rotation offers a viable strategy for investigating the non-perturbative properties of a rotating gluon system. However, this generalization is actually not so straightforward as it sounds due to the inherent inhomogeneity of a rotating system.

The inhomogeneous problem in rotating matter have attracted great interests in different branches of physics, such as the vortex state of rotating cold atom systems~\cite{Fetter:2009zz, Matthews:1999zz, Madison:2000zz, Haljan:2001zz} and the glitch behavior of pulsar~\cite{Melatos:2007px, Chamel:2012ae}. For the many-quark system inhomogeneous chiral condensate has also been studied in~\cite{Jiang:2016wvv} in local potential approximation and in~\cite{Wang:2018zrn} self-consistently with a modified BdG equation. For gluon system and deconfinement problem, the mechanism of color confinement is no longer particle pairing, thus a self-consistent computation of soliton spatial distribution should be considered. With such kind of effective models and lattice QCD simulation, several great works have been done by introducing the rotation with the rotwisted boundary condition~\cite{Chen:2020ath, Fujimoto:2021xix, Chernodub:2020qah, Chernodub:2022wsw, Chen:2022smf, Chernodub:2022veq, Braguta:2023iyx, Braguta:2021ucr, Braguta:2021jgn, Yang:2023vsw}. In this work we will start from the SU(2) Yang-Mills theory and solve dyons under rotation with the usual periodic boundary condition. In our last work~\cite{Jiang:2023zzu} the on-axis dyon solutions have been obtained. Such solutions can only be used to study the system in a cell on axis. In order to study the inhomogeneity of the system we will solve them at arbitrary position because the rotation-modified Yang-Mills equation depends on the radial coordinates explicitly. 

\begin{figure}[ht]
    \centering
    \includegraphics[width=6 cm]{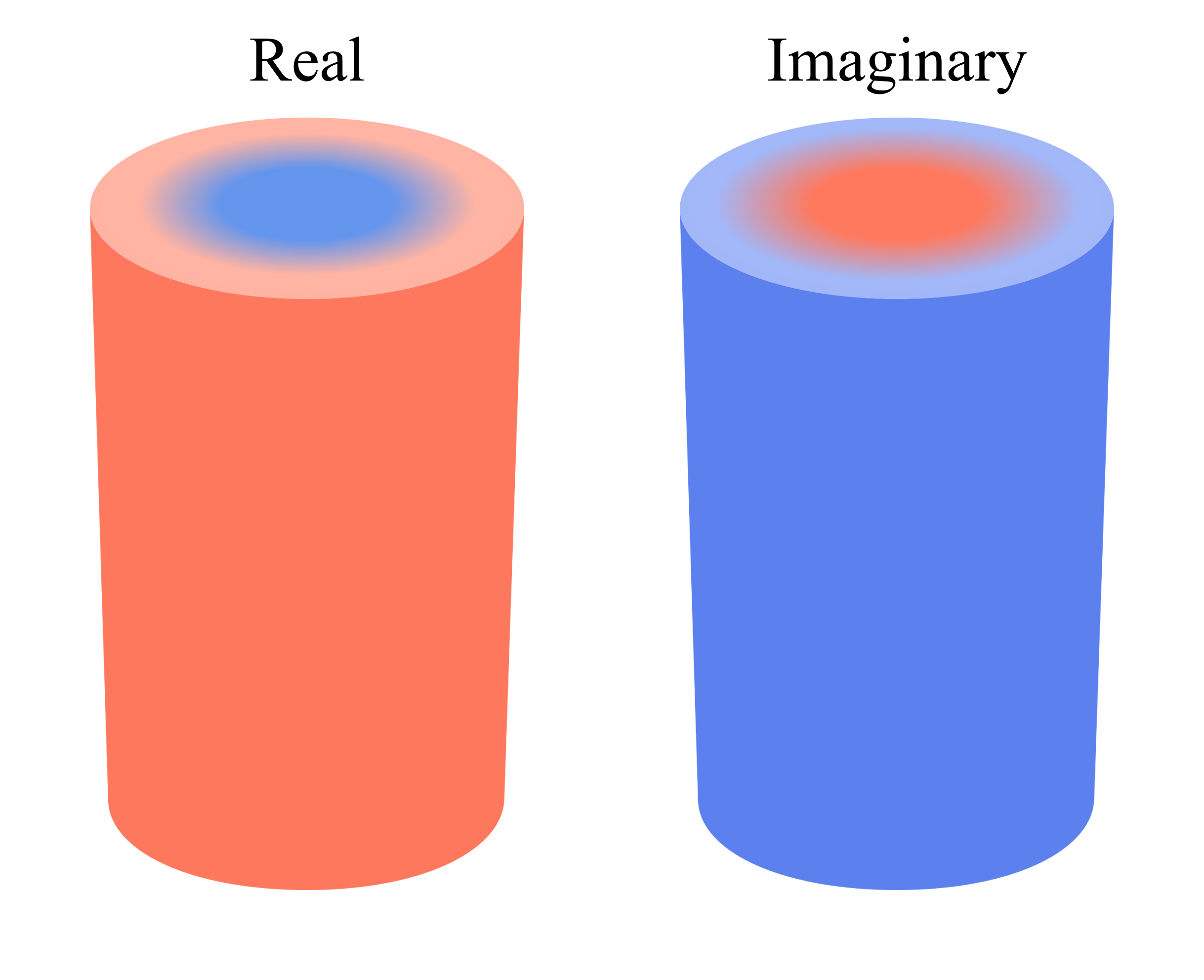}
    \caption{A schematic illustration of the radial position dependence of the deconfinement temperature in both real(left) and imaginary(right) angular velocity cases.}
    \label{fig0}
\end{figure}

The inhomogeneity problem in rotating matter has been studied across various domains of physics, such as the vortex state of rotating cold atom systems and the glitch behavior observed in pulsars. Within the realm of many-quark systems, the inhomogeneous chiral condensate has been explored using both local potential approximations and a self-consistent approach incorporating a modified BdG equation. Regarding the gluon system and the deconfinement problem, the underlying mechanism is no longer particle pairing, thus a self-consistent computation to ascertain the soliton spatial distribution is more suitable. Leveraging such effective models of solitons and lattice QCD simulations with imaginary velocity, pioneering research has been undertaken by incorporating rotation through the implementation of rotwisted boundary conditions. In the present study, we will focus on the SU(2) Yang-Mills theory to resolve dyons under rotation with the standard periodic boundary conditions. In our preceding work, strict solutions for dyons on the spinning axis were successfully derived. However, these solutions are exclusively applicable for examining the thermodynamics within an on-axis cell. To delve into the system's inhomogeneity, we are committed to resolving the rotation-modified Yang-Mills equation exhibiting an explicit dependence on radial coordinates. And it will be found that in the real velocity case the outer layer will deconfine more difficult while the reverse is true in imaginary velocity case as shown in the schematic picture in Fig.~\ref{fig0}. 

\section{Strict dyon solutions in static and spinning cases}
In this section we briefly review strict BPS dyon solutions in the static case first and then the spinning case. Using these semi-classical solitons one can model the confinement-deconfinement transition of a SU(2) gluon system at finite temperature successfully. There are two types of dyons in the SU(2) Yang-Mills theory, i.e. M and L dyons and their anti-dyons $\bar{\text M}$ and $\bar{\text L}$. And in the hedge-hog gauge a single M(+) or $\bar{\text M}$(-), whose center locates at $\vec{c}$, is
\begin{eqnarray}
\label{mdyon}
&&A^a_4(x; \rho)=\pm n_a (\frac{1}{r}-\rho\ \coth(\rho r))\\
&&A^a_m(x; \rho)=\epsilon_{a m k}n_k(\frac{1}{r}-\rho\ \text{csch}(\rho r)).
\end{eqnarray}
where $\vec{r}=\vec{x}-\vec{c}$ and $\vec{n}=\vec{r}/r$. And L and $\bar{\text L}$ dyons are obtained by replacing the parameter $\rho$ with $\rho_c=2\pi T-\rho$. Intuitively this parameter $\rho$ characterizes the size of a dyon. In stringy gauge all the dyons will be gauge transformed to have the same asymptotic value at spatial infinity. They are strict solutions of the Yang-Mills equation in the Euclidean time-space ($x_4=i t$) which corresponding to a many-gluon system at finite temperature
\begin{flalign}
\label{ym0}
&& &\partial_i G_{4i}^a=-\epsilon_{a b c}G_{4 i}^c A_i^b&\\
&& &\partial_i G_{ij}^a=\epsilon_{a b c}(G_{4 j}^b A_4^c+G_{j n}^c A_n^b).&\nonumber
\end{flalign}
In such a imaginary-time thermal field theory, BPS dyons are good solutions because they satisfy the periodic boundary condition along the imaginary temporal direction. And in static case the dyon and anti-dyon solutions are actually sef-dual and anti-self-dual by considering the Bogomol’nyi inequality~\cite{Diakonov:2004jn}. Dyon solutions are non-trivial because their temporal components $A^a_4\rightarrow \mp\rho n_a$ can be nonzero at spatial infinity, which leads to a nonzero trace of the Polyakov loop
\begin{eqnarray}
\label{pl}
Tr(L)=Tr(\mathcal{P} e^{i \int_0^\beta dx_4 A_4(x)})=2\cos(\beta\rho/2),
\end{eqnarray}
at high temperature(deconfinement) and approaching zero(confinement) smoothly as the fall of temperature. It means that the parameter $\rho$ can be treated as an equivalent order parameter of confinement. And the $\rho$'s value is usually solved self-consistently by minimizing the effective potential of the system. Even when the $\rho$ is fixed there are still infinite number of dyon solutions because the Eq.~\ref{ym0} is translational invariant. This means an arbitrary vector $\vec{c}$ gives a different solution of dyon. 

To construct a SU(2) gluon system with the semi-classical degree of freedom, i.e. dyon, one need to obtain the dyon-dyon and dyon-anti-dyon interaction by solving the semi-classical solution with larger topological charge. These solutions can be treated as superposition of multi-dyons or anti-dyons in a specific way, i.e. the Atiyah--Hitchin--Drinfeld--Manin (AHDM) construction~\cite{Diakonov:2004jn}. By separating the effective action of such a solutions into summation of single dyons and residual contributions, the residual terms can be identified as the interaction between dyons/anti-dyons~
\cite{Diakonov:2005qa}. Once a type of non-trivial configurations $\bar A(x; \{c_i\})$ for the vacuum has been chosen, the 1-loop partition function of the system is calculated as the standard formalism
\begin{eqnarray}
\label{za}
    &&\mathcal{Z}=\int D A_\mu \exp(-S[A])\nonumber\\
    &&=\int \Pi_{i}d\xi_i \exp(-S[\bar A])Det(-D^2)
\end{eqnarray}
where $S[A]=\int d^4x G^2/4g^2$ for a general Yang-Mills field. 
In the dilute limit the interaction is negligible and the $e^{-S[\bar A]}$ can be computed by summing one L/$\bar{\text L}$ and one M/$\bar{\text M}$ dyon's action together. Consider the vacuum can be a more general system composed with more than one pair of L/$\bar{\text L}$ and M/$\bar{\text M}$, such a dyon ensemble gives the partition function as
\begin{eqnarray}
\label{zfunc}
    &&\mathcal{Z}_D=\sum_{N_M, N_{\bar M}, N_L, N_{\bar L}}\frac{1}{N_M!N_{\bar M}!N_L!N_{\bar L}!}\nonumber\\
     &&\times\left[\int d^3c_M |{\bar\rho}|^3 \exp(-S[A_M])\right]^{N_M+N_{\bar M}}\nonumber\\
    &&\times\left[\int d^3c_L |{\bar\rho_c}|^3 \exp(-S[ A_L])\right]^{N_L+N_{\bar L}}
\end{eqnarray}
where $\bar\rho$ is the uniform asymptotic value for both L and M dyons in stringy gauge and $\rho d^3c$ serves as the integral measure. As a result the non-perturbative semi-classical potential in dilute limit is
\begin{eqnarray}
    &&U_{np}=-\frac{1}{V}ln(\mathcal{Z}_D)\nonumber\\
    &&=-c\left[{|\bar\rho|}^3 (\frac{\Lambda}{\pi T})^{\frac{22\bar\rho}{6\pi T}}
    +|{\bar\rho}_c|^3 (\frac{\Lambda}{\pi T})^{\frac{22|\bar\rho_c|}{6\pi T}}\right]
\end{eqnarray}
where $\bar\rho_c=2\pi T-\bar\rho$. Besides doing the full quantum weight computation in Eq.~\ref{za}, the perturbative part can also be obtained by setting the an uniform background $A_4=\bar\rho \tau_3/2$ and integrating out fluctuations of plane waves about it. It has been done as~\cite{Gross:1980br} and gives the static perturbative potential as
\begin{eqnarray}
    &&U_p=\frac{1}{12\pi^2 T}\bar\rho^2{\bar\rho_c}^2
\end{eqnarray}
The order parameter $\bar\rho$ value is determined by minimizing the total effective potential $U_{np}+U_p$ at different temperature. 
With typical parameters one can solve that $\bar\rho=\pi T$ corresponding to confinement at low temperature and deviates from $\pi T$ which means deconfinement as temperature increase.

From the very brief review in above one can find that the explicit semi-classical solution is the key quantity for modelling the deconfinement phase transition. Once a category of soliton has been obtained the effective potential can be computed according to Eq.~\ref{zfunc} by considering a classical soliton ensemble. For a rotating SU(2) gluon system with uniform angular velocity $i \vec\omega$, the task is to find a solution like   
\begin{flalign}
\label{rot}
&& &A=A(x_4, \varrho, \phi+\omega x_4, z),&
\end{flalign}
satisfying the usual SU(2) Yang-Mills equation. The {\it centripetal force} holding the rotation of each cell in the system is provided by the fundamental interaction of gluons. The rotwisted boundary condition~\cite{Chernodub:2020qah} along the $x_4$ dimension requires
\begin{flalign}
&& &A(x_4+\beta, \varrho, \phi-\beta\omega+\omega (x_4+\beta), z)&\nonumber\\
&& &=A(x_4, \varrho, \phi+\omega x_4, z).&
\end{flalign}
It is widely applied in the lattice formalism because no modification need to be introduced into the Lagrangian density besides considering a unusual boundary condition of each gauge field configuration. But it is not quite convenient for analytical computation. By noticing the time derivative relation as follows
\begin{eqnarray}
\label{pd0}
\partial_4 A&=&A^{(1, 0, 0, 0)}(x_4, \varrho, \phi+\omega x_4, z)\nonumber\\
&+&\omega A^{(0, 0, 1, 0)}(x_4, \varrho, \phi+\omega x_4, z),
\end{eqnarray}
one can find a equivalent formalism for this problem as follows. By solving a rotation-modified Yang-Mills equation in Eq.~\ref{rotym0} with the usual periodic boundary condition, 
the Yang-Mills equation is
\begin{flalign}
\label{rotym0}
&\partial_i G_{4i}^a=-\epsilon_{a b c}G_{4 i}^c A_i^b&\\
&\partial_i G_{ij}^a=v_i\partial_i G_{4j}^a+\partial_j v_i G_{4 i}^a+\epsilon_{a b c}(G_{4 j}^b A_4^c+G_{j n}^c A_n^b)&\nonumber
\end{flalign}
where the field strength tensor is 
\begin{flalign}
&G_{4 i}^a=\partial_4 A_i^a-\partial_i A_4^a+\epsilon_{a b c}A_4^b A_i^c+\partial_j v_i A_j^a-v_j \partial_j A_i^a&\nonumber\\
&G_{i j}^a=\partial_i A_j^a-\partial_j A_i^a+\epsilon_{a b c}A_i^b A_j^c&
\end{flalign}
one can get a solution satisfying $A(x_4, \varrho, \phi, z)=A(x_4+\beta, \varrho, \phi, z)$. Then the target solution in Eq.~\ref{rot} can been obtained as well by simply replacing the $\phi$ here with $\phi+\omega x_4$. Evidently this procedure corresponds to going into a co-rotating frame in which a curved metrics will produce the extra terms from the derivative in Eq.~\ref{pd0}. In the following we will follow this approach to take the rotation into account.

In the rotating case with an imaginary angular velocity, one can not expect a spherical symmetric solution as Eq.~\ref{mdyon} because of the cylindrical symmetry of the system. However, when dyons locate at the rotation axis $c_{1, 2}=0$, the rotation correction is very simple and strict. In our last work~\cite{Jiang:2023zzu} it has been found that such a on-axis solution $A^{D, rot}(x)$ is like
\begin{eqnarray}
A^{D, rot}=(A^{D}_4(x; \rho_D)+\delta A_4, ~ \vec{A}^{D}(x;\rho_D)).
\end{eqnarray}
where the $\rho_D$ is the parameter in the static solution of dyon $A^{D}(x; \rho)$ and the correction under rotation in the hedge-hog gauge is 
\begin{eqnarray}
\label{oax}
&&\delta A_4=\mp \omega_a \frac{\tau^a}{2}.
\end{eqnarray}
where $\mp$ for the M and L dyon and $\pm$ for ${\bar{\text M}}$ and ${\bar{\text L}}$ respectively and $\tau_a, a=1, 2, 3$ are Pauli matrices. Clearly the asymptotic value of $A_4$ has been shifted by $\pm \omega$ for different kind of dyons, but it can be checked explicitly that the solution will not change the field strength tensor or dyon action. By requiring the potential multiple dyon/anti-dyon solution has a uniform asymptotic behavior $\bar\rho$ at spatial infinity, the parameter $\rho_D$ in the solution should be chosen to cancel the $\omega$-dependent terms. And this makes the action of each constituent dyon becomes
\begin{eqnarray}
    &&S_{M, \bar M}=\frac{(\bar\rho \pm \omega)}{2\pi T}\frac{8\pi^2}{g^2}\nonumber\\
    &&S_{L, \bar L}=\frac{(\bar\rho_c \pm \omega)}{2\pi T}\frac{8\pi^2}{g^2}
\end{eqnarray}
where $\bar\rho_c=2\pi T-\bar\rho$ and the scale of this model  $\Lambda$ is related with the running coupling as $(\frac{\Lambda}{\pi T})^{22/3}=e^{-8\pi^2/g^2}$. Considering the non-interacting dyon ensemble and noticing the dyon and anti-dyon actions are now different, the partition function Eq.~\ref{zfunc} should be changed as
\begin{eqnarray}
\label{zfunc2}
    &&\mathcal{Z}_{rot, D}=\sum_{N_M, N_{\bar M}, N_L, N_{\bar L}}\frac{1}{N_M!N_{\bar M}!N_L!N_{\bar L}!}\nonumber\\
     &&\times\left[\int d^3c_M sgn(\bar\rho)(\bar\rho+\omega)^3 \exp(-S[\bar A_M])\right]^{N_M}\nonumber\\
     &&\times\left[\int d^3c_{\bar M} sgn(\bar\rho)(\bar\rho-\omega)^3 \exp(-S[\bar A_{\bar M}])\right]^{N_{\bar M}}\nonumber\\
    &&\times\left[\int d^3c_L sgn(\bar\rho_c)(\bar\rho_c+\omega)^3 \exp(-S[\bar A_L])\right]^{N_L}\nonumber\\
     &&\times\left[\int d^3c_{\bar L} sgn(\bar\rho_c)(\bar\rho_c-\omega)^3 \exp(-S[\bar A_{\bar L}])\right]^{N_{\bar L}}
\end{eqnarray}
The $sgn(\rho)$ here corresponding to the $|\rho|$ in Eq.~\ref{zfunc}, is from that the $\rho\tau_3/2=diag(\rho/2, -\rho/2)$ can be always replaced with $diag(sgn(Re(\rho))\rho/2, -sgn(Re(\rho))\rho/2)$ in the action computation by a global gauge transformation. Completing the summation, the non-perturbative potential is extracted as 
\begin{eqnarray}
&U_{np}(T, \omega)=-\frac{1}{V}ln \mathcal{Z}_{rot, D}\nonumber\\
&=-\frac{c}{2}[sgn(\bar\rho)(\bar\rho+\omega)^3(\frac{\Lambda}{\pi T})^{\frac{22sgn(\bar\rho)(\bar\rho+\omega)}{6\pi T}}\nonumber\\
&+sgn(\bar\rho_c)(\bar\rho_c+\omega)^3(\frac{\Lambda}{\pi T})^{\frac{22sgn(\bar\rho_c)(\bar\rho_c+\omega)}{6\pi T}}]\nonumber\\
&-\frac{c}{2}[sgn(\bar\rho)(\bar\rho-\omega)^3(\frac{\Lambda}{\pi T})^{\frac{22sgn(\bar\rho)(\bar\rho-\omega)}{6\pi T}}\nonumber\\
&+sgn(\bar\rho_c)(\bar\rho_c-\omega)^3(\frac{\Lambda}{\pi T})^{\frac{22sgn(\bar\rho_c)(\bar\rho_c-\omega)}{6\pi T}}].
\end{eqnarray}
This difference between caloron $\bar\rho$ and anti-caloron $\bar\rho_c$ contributions is induced by the parity violation in a globally rotating system and will vanish when $\omega$ goes to zero.

The perturbative potential $U_p$ are obtained by integrating out Gaussian fluctuations on top of a multi-dyon background according to reference~\cite{Gross:1980br}. In such a rotating system eigen states of quantum fluctuations around the semi-classical gauge field are cylindrical waves involving with Bessel functions. The momentum integration is another three good quantum numbers, i.e. transverse momentum $k_T=\xi_n^{(m)}/R$, longitudinal momentum $k_z$ and angular momentum number $m$ corresponding to the polar radius, longitudinal coordinate and azimuthal angle, where $R$ is the size of system and $\xi_n^{(m)}$ the m-th zero of the Bessel function $J_n(r)$. The transverse momentum is discretized by considering the finite boundary condition $J_n(k_T R)=0$. The whole computation procedure is straightforward but a little lengthy for this paper. One can refer Ref.~\cite{Fujimoto:2021xix} for more details. The result is the same as the reference 
\begin{eqnarray}
\label{pert}
U_{p}(T, \omega)=&-\sum\limits_{\substack{s, m=1\\n=-\infty}}^{+\infty}\frac{e^{\frac{i s n \omega}{T} }cosh(\frac{i s \omega}{T})}{\pi^2 s R^3}\frac{4 \xi_n^{(m)}cos(s\frac{\bar\rho}{T})}{J_{n+1}(\xi_n^{(m)})^2}\nonumber\\
&\times J_{n}(\xi_n^{(m)}\frac{r}{R})^2 K_1(s \frac{\xi_n^{(m)}}{T R}).
\end{eqnarray}
where $r=0$ for the on-axis case. One should beware that in all the above computation the angular velocity is actually an imaginary one. And the result for the real velocity case can be easily obtained by replacing the $\omega$ with $i\omega$. Although the rotation correction is strict and simple for on-axis dyons, it is not enough for us to explore the radial dependence of a rotating system. To study such a inhomogeneous property one should find rotation corrections to dyons with arbitrary centers.

\section{Rotation-correction to off-axis dyons}
In this section we solve the off-axis dyon corrections with the imaginary angular velocity. One will find the whole procedure is straightforward algebraic computation, so the real velocity results will be obtained by replacing $\omega$ with $i\omega$ eventually. As analyzed from the aspect of symmetry in last section, it is natural that for a off-axis dyon the rotation-induced correction are not so simple and strict as the on-axis case in Eq.~\ref{oax}. Hence we expand the solution as series of $\omega^n$ or more precisely in powers of velocity multiplications $\vec{v}^n\vec{u}^m$
\begin{flalign}
A=A^{(0)}(\omega^0)+A^{(1)}(\omega^1)+A^{(2)}(\omega^2)+...
\end{flalign}
where $\vec{u}=\vec{\omega}\times\vec{c}$
and solve the Eq.~\ref{rotym0} to the $\mathcal{O}(\omega^2)$.
When the $\mathcal{O}(\omega^0)$ solution is a off-axis dyon at $\vec{c}=(y_1, y_2, y_3)$, it is more convenient to redefine the $\vec{x}-\vec{c}$ as the new $\vec{x}$. Such a translation can preserve the spherical symmetry of the $\omega^0$ solution which is better for the following analytical derivation. In the new $\vec{x}$ coordinate the Yang-Mills equation is
\begin{flalign}
\label{rotym}
&\partial_i G_{4i}^a=-\epsilon_{a b c}G_{4 i}^c A_i^b&\\
&\partial_i G_{ij}^a=(v_i+u_i)\partial_i G_{4j}^a+\partial_j v_i G_{4 i}^a+\epsilon_{a b c}(G_{4 j}^b A_4^c+G_{j n}^c A_n^b)&\nonumber
\end{flalign}
where the field strength tensor is 
\begin{flalign}
&G_{4 i}^a=\partial_4 A_i^a-\partial_i A_4^a+\epsilon_{a b c}A_4^b A_i^c+\partial_j v_i A_j^a-(v_j+u_j) \partial_j A_i^a&\nonumber\\
&G_{i j}^a=\partial_i A_j^a-\partial_j A_i^a+\epsilon_{a b c}A_i^b A_j^c&
\end{flalign}
and $\vec{u}$ is from the shift of coordinates. Noting that that both $\vec{v}$ and $\vec{u}$ are $\omega^1$ order, expanding these equations to $\omega^2$ gives
\begin{flalign}
&\partial_i G_{4i}^{(0)a}=-\epsilon_{a b c}G_{4 i}^{(0)c} A_i^{(0)b}&\\
&\partial_i G_{ij}^{(0)a}=\epsilon_{a b c}(G_{4 j}^{(0)b} A_4^{(0)c}+G_{j n}^{(0)c} A_n^{(0)b})&\nonumber
\end{flalign}
as the equations in static case and the $\mathcal{O}(\omega^1)$ and $\mathcal{O}(\omega^2)$ equations are
\begin{flalign}
\label{o1}
&\partial_i G_{4i}^{(1)a}=-\epsilon_{a b c}(G_{4 i}^{(0)c} A_i^{(1)b}+G_{4 i}^{(1)c} A_i^{(0)b})&\\
&\partial_i G_{ij}^{(1)a}=(v_i+u_i)\partial_i G_{4j}^{(0)a}+\partial_j v_i G_{4 i}^{(0)a}&\nonumber\\
&+\epsilon_{a b c}(G_{4 j}^{(0)b} A_4^{(1)c}+G_{4 j}^{(1)b} A_4^{(0)c}+G_{j n}^{(0)c} A_n^{(1)b}+G_{j n}^{(1)c} A_n^{(0)b})&\nonumber
\end{flalign}
and
\begin{flalign}
\label{o2}
&\partial_i G_{4i}^{(2)a}=-\epsilon_{a b c}(G_{4 i}^{(1)c} A_i^{(1)b}+G_{4 i}^{(0)c} A_i^{(2)b}+G_{4 i}^{(2)c} A_i^{(0)b})&\nonumber\\
&\partial_i G_{ij}^{(2)a}=(v_i+u_i)\partial_i G_{4j}^{(1)a}+\partial_j v_i G_{4 i}^{(1)a}&\nonumber\\
&+\epsilon_{a b c}(G_{4 j}^{(1)b} A_4^{(1)c}+G_{4 j}^{(0)b} A_4^{(2)c}+G_{4 j}^{(2)b} A_4^{(0)c}&\nonumber\\
&+G_{j n}^{(1)c} A_n^{(1)b}+G_{j n}^{(0)c} A_n^{(2)b}+G_{j n}^{(2)c} A_n^{(0)b})&
\end{flalign}
where the $\omega$-expansion of the field strength tensor is
\begin{flalign}
&G_{4 i}^{(0)a}=\partial_4 A_i^{(0)a}-\partial_i A_4^{(0)a}+\epsilon_{a b c}A_4^{(0)b} A_i^{(0)c}&\nonumber\\
&G_{i j}^{(0)a}=\partial_i A_j^{(0)a}-\partial_j A_i^{(0)a}+\epsilon_{a b c}A_i^{(0)b} A_j^{(0)c}&
\end{flalign}
as the $\mathcal{O}(\omega^0)$ order and 
\begin{flalign}
&G_{4 i}^{(1)a}=\partial_4 A_i^{(1)a}-\partial_i A_4^{(1)a}
+\partial_j v_i A_j^{(0)a}-(v_j+u_j) \partial_j A_i^{(0)a}&\nonumber\\
&+\epsilon_{a b c}(A_4^{(0)b} A_i^{(1)c}+A_4^{(1)b} A_i^{(0)c})&\\
&G_{i j}^{(1)a}=\partial_i A_j^{(1)a}-\partial_j A_i^{(1)a}+\epsilon_{a b c}(A_i^{(1)b} A_j^{(0)c}+A_i^{(0)b} A_j^{(1)c})&\nonumber
\end{flalign}
and 
\begin{flalign}
&G_{4 i}^{(2)a}=\partial_4 A_i^{(2)a}-\partial_i A_4^{(2)a}
+\partial_j v_i A_j^{(1)a}-(v_j+u_j) \partial_j A_i^{(1)a}&\nonumber\\
&+\epsilon_{a b c}(A_4^{(1)b} A_i^{(1)c}+A_4^{(0)b} A_i^{(2)c}+A_4^{(2)b} A_i^{(0)c})&\nonumber\\
&G_{i j}^{(2)a}=\partial_i A_j^{(2)a}-\partial_j A_i^{(2)a}&\nonumber\\
&+\epsilon_{a b c}(A_i^{(1)b} A_j^{(1)c}+A_i^{(2)b} A_j^{(0)c}+A_i^{(0)b} A_j^{(2)c})&
\end{flalign}
the $\mathcal{O}(\omega^1)$ and $\mathcal{O}(\omega^2)$ terms. Taking the $\bar M$ dyon for example, after substituting its spherical symmetric solution as the $A^{(0)a}_\mu$ into above equations, although there are many terms in each expression, Eq.~\ref{o1} and \ref{o2} are actually all non-homogeneous linear equations of $A^{(1)a}_\mu$ and $A^{(2)a}_\mu$. This means these equations are solvable in principle with the help of corresponding Green functions. However, the straightforward computation is too tedious because all the components of field strength tensor are coupling together. Instead, we can try to find the tensor structure by noticing non-homogeneous terms. Taking the $G_{4 i}^a$ equations for example. Non-homogeneous terms in Eq.~\ref{o1} are in the R.H.S of the following equation
\begin{flalign}
\mathcal{L}^a[A_4^{(1)}, A_k^{(1)}]=(-x_m \omega_m x_a-\omega_a r^2+\epsilon_{a i j}u_i x_j)Q^2(r)
\end{flalign}
where $\mathcal{L}^a$ is the homogeneous part in Eq.~\ref{o1} and $r^2=x_m x_m$. Then we can factorize $A_4^{(1)}$ according to the non-homogeneous term as
\begin{flalign}
A_4^{(1)a}=R_1(r)\omega_a+R_2(r)x_m \omega_m x_a+R_3(r)\epsilon_{a i j}u_i x_j
\end{flalign}
The first terms gives $\omega_a$ in the on-axis solution. And analyzing the $A_i^{(1)a}$ equation in the same way one can choose
\begin{flalign}
A_i^{(1)a}=S_1(r)u_i x_a+S_2(r)u_a x_i+(S_3(r) x_i x_a+S_4(r)\delta_{i a})u_m x_m
\end{flalign}
Substituting the two factorizations into Eq.~\ref{o1} we can obtain the solutions
\begin{flalign}
&& &A_4^{(1)a}=\omega_a-\epsilon_{i j a}u_i x_j Q(r)&\nonumber\\
&& &A_i^{(1)a}=u_i x_a P(r)&
\end{flalign}
Once the $\mathcal{O}(\omega^1)$ solutions have been obtained, the Eq.~\ref{o2} can be solved with the same procedure. The corresponding non-homogeneous terms indicate second order corrections can be factorized as
\begin{flalign}
&A_4^{(2)a}={\Tilde{R}}_1 u^2 x_a+{\Tilde{R}}_2 (u_m x_m)^2 x_a+{\Tilde{R}}_3 u_m x_m u_a&\nonumber\\
&+{\Tilde{R}}_4 \xi_a+{\Tilde{R}}_5 \xi_m x_m x_a&\nonumber\\
&A_i^{(2)a}={\Tilde{S}}_1(\omega_m u_i x_m x_a-\omega_i u_m x_m x_a)+{\Tilde{S}}_2\epsilon_{a m n}u_m x_n u_i&\nonumber\\
&+{\Tilde{S}}_3\epsilon_{i m n}u_m x_n u_a+{\Tilde{S}}_4 u_k x_k \epsilon_{a m n}u_m x_n x_i&\nonumber\\
&+{\Tilde{S}}_5 u_k x_k \epsilon_{i m n}u_m x_n x_a+{\Tilde{S}}_6 u^2 \epsilon_{i a m}x_m&\nonumber\\
&+{\Tilde{S}}_7 u_k x_k \epsilon_{i a m}u_m+{\Tilde{S}}_8 (u_k x_k)^2 \epsilon_{i a m}x_m,&
\end{flalign}
where $\vec{\xi}=\vec{\omega}\times\vec{u}$.
Substituting them into Eq.~\ref{o2}, we obtain the second order corrections as 
\begin{flalign}
&A_4^{(2)a}=(Q(r)-\frac{1}{2}r^2 Q^2(r)) u^2 x_a+\frac{1}{2}(-1+r^2 Q(r)) \xi_a&\nonumber\\
&-\frac{1}{2r^2}(P(r)+2Q(r)-r^2 Q^2(r)) (u_m x_m)^2 x_a&\nonumber\\
&+\frac{1}{2}P(r) u_m x_m u_a+\frac{1}{2}Q(r) \xi_m x_m x_a&\nonumber\\
&A_i^{(2)a}=\frac{1}{2}P(r)(\omega_m u_i x_m x_a-\omega_i u_m x_m x_a)+\frac{1}{2r^2}\epsilon_{a m n}u_m x_n u_i&\nonumber\\
&+\frac{1}{2}(-Q(r)+\frac{1}{r^2})\epsilon_{i m n}u_m x_n u_a-(Q(r)-\frac{3}{2r^2}) u_k x_k \epsilon_{i a m}u_m&\nonumber\\
&+\frac{1}{r^2}(-\frac{2}{r^2}+Q(r)) u_k x_k \epsilon_{a m n}u_m x_n x_i&\nonumber\\
&-\frac{1}{2}(P(r)-Q(r)-r^2 P(r)Q(r)+\frac{1}{r^2}) u^2 \epsilon_{i a m}x_m&\nonumber\\
&-\frac{1}{2r^2}(-P(r)+Q(r)+r^2 P(r)) (u_k x_k)^2 \epsilon_{i a m}x_m.&
\end{flalign}
Now we can obtain the final solution by shifting $\vec{x}$ back to $\vec{x}-\vec{c}$. Then substituting the solution into the Yang-Mills action and complete the integration we obtain the action of the off-axis dyon with the center at $\vec{c}=(y_1, y2, 0)$ as
\begin{flalign}
S_{\bar M}=\frac{8\pi^2}{g^2}\frac{\rho_{\bar M}}{2\pi T}(1+\omega^2 (y_1^2+y_2^2))
\end{flalign}
The difference between $\mathcal{O}(\omega^0)$ solutions of $\text{M}$ and $\bar{\text{M}}$ dyon is the sign of the spatial part, which means the higher order correction of $\text{M}$ dyon is 
\begin{flalign}
&& &A_4^{(1)}[\text{M}]=A_4^{(1)}[\bar{\text{M}}]&\nonumber\\
&& &A_i^{(1)}[\text{M}]=-A_i^{(1)}[\bar{\text{M}}]&\nonumber\\
&& &A_4^{(2)}[\text{M}]=-A_4^{(2)}[\bar{\text{M}}]&\nonumber\\
&& &A_i^{(2)}[\text{M}]=A_i^{(2)}[\bar{\text{M}}]&
\end{flalign}

Because all the above derivation is in the hedgehog gauge, the $\text{L}$ and $\bar{\text{L}}$ dyons can be obtained by replacing the $\rho_D$ with $2\pi T-\rho_D$ and applying gauge transformation to the stringy gauge. As the gauge transformation does not change action, the action of all kinds of dyons in the imaginary velocity case are
\begin{flalign}
S_{D}=\frac{8\pi^2}{g^2}\frac{\rho_{D}}{2\pi T}(1+\omega^2 (y_1^2+y_2^2))
\end{flalign}
To construct a multi-dyon solution with an uniform asymptotic behavior, the $\rho_D$ are $\bar\rho \mp \omega$ and $2\pi T-\bar\rho \pm \omega$, where $\mp$ and $\pm$ corresponding to $\text{M}$, $\bar{\text{M}}$, $\text{L}$ and $\bar{\text{L}}$ dyons respectively. As a result, the non-perturbative potential is 
\begin{flalign}
&& &U^{RE}_{np}(T, \omega)=-\frac{1}{V}ln \mathcal{Z}_{rot, D}&\nonumber\\
&& &=-\frac{c}{2}\alpha^3(\omega)[sgn(\bar\rho)(\bar\rho+i \omega)^3(\frac{\Lambda}{\pi T})^{\frac{22sgn(\bar\rho)\alpha(\bar\rho+i\omega)}{6\pi T}}&\nonumber\\
&& &+sgn(\bar\rho_c)(\bar\rho_c+i\omega)^3(\frac{\Lambda}{\pi T})^{\frac{22sgn(\bar\rho_c)\alpha(\bar\rho_c+i\omega)}{6\pi T}}]&\nonumber\\
&& &-\frac{c}{2}\alpha^3(\omega)[sgn(\bar\rho)(\bar\rho-i\omega)^3(\frac{\Lambda}{\pi T})^{\frac{22sgn(\bar\rho)\alpha(\bar\rho-i\omega)}{6\pi T}}&\nonumber\\
&& &+sgn(\bar\rho_c)(\bar\rho_c-i\omega)^3(\frac{\Lambda}{\pi T})^{\frac{22sgn(\bar\rho_c)\alpha(\bar\rho_c-i\omega)}{6\pi T}}].&
\end{flalign}
where $\alpha=1-\omega^2(y_1^2+y_2^2)$ in the real rotation velocity case. And 
\begin{flalign}
&& &U^{IM}_{np}(T, \omega)=-\frac{1}{V}ln \mathcal{Z}_{rot, D}&\nonumber\\
&& &=-\frac{c}{2}\alpha^3(\omega)[sgn(\bar\rho)(\bar\rho+ \omega)^3(\frac{\Lambda}{\pi T})^{\frac{22sgn(\bar\rho)\alpha(\bar\rho+\omega)}{6\pi T}}&\nonumber\\
&& &+sgn(\bar\rho_c)(\bar\rho_c+\omega)^3(\frac{\Lambda}{\pi T})^{\frac{22sgn(\bar\rho_c)\alpha(\bar\rho_c+\omega)}{6\pi T}}]&\nonumber\\
&& &-\frac{c}{2}\alpha^3(\omega)[sgn(\bar\rho)(\bar\rho-\omega)^3(\frac{\Lambda}{\pi T})^{\frac{22sgn(\bar\rho)\alpha(\bar\rho-\omega)}{6\pi T}}&\nonumber\\
&& &+sgn(\bar\rho_c)(\bar\rho_c-\omega)^3(\frac{\Lambda}{\pi T})^{\frac{22sgn(\bar\rho_c)\alpha(\bar\rho_c-\omega)}{6\pi T}}].&
\end{flalign}
where $\alpha=1+\omega^2(y_1^2+y_2^2)$ in the imaginary rotation velocity case.

In the off-axis case we adopt the perturbative potential $U_p$ as Eq.~\ref{pert} in last section because we only consider the lowest order correction induced by the background rotation which is supposed to be a even function of of $\omega$, i.e. $\mathcal{O}(\omega^2)$. The results in the real and imaginary angular velocity case are respectively
\begin{eqnarray}
\label{pert}
U^{RE}_{p}(T, \omega)=&-\sum\limits_{\substack{s, m=1\\n=-\infty}}^{+\infty}\frac{e^{\frac{s n \omega}{T} }cosh(\frac{s \omega}{T})}{\pi^2 s R^3}\frac{4 \xi_n^{(m)}cos(s\frac{\bar\rho}{T})}{J_{n+1}(\xi_n^{(m)})^2}\nonumber\\
&\times J_{n}(\xi_n^{(m)}\frac{r_c}{R})^2 K_1(s \frac{\xi_n^{(m)}}{T R}).
\end{eqnarray}
and
\begin{eqnarray}
\label{pert}
U^{IM}_{p}(T, \omega)=&-\sum\limits_{\substack{s, m=1\\n=-\infty}}^{+\infty}\frac{e^{\frac{i s n \omega}{T} }cosh(\frac{i s \omega}{T})}{\pi^2 s R^3}\frac{4 \xi_n^{(m)}cos(s\frac{\bar\rho}{T})}{J_{n+1}(\xi_n^{(m)})^2}\nonumber\\
&\times J_{n}(\xi_n^{(m)}\frac{r_c}{R})^2 K_1(s \frac{\xi_n^{(m)}}{T R}).
\end{eqnarray}
where $r_c=\sqrt{y_1^2+y_2^2}$ for the off-aixs case.
Clearly in both non-perturbative and perturbative computation we have only kept the lowest order correction, i.e. $\mathcal{O}(\omega^2)$ and there is no $\mathcal{O}(\omega)$ term in the effective potential. It is reasonable because the $\vec\omega$ direction is supposed to be non-relevant to the thermodynamics of the system. In the real velocity case, the negative sign in $\alpha$ requires the $|\vec{u}|$ can not be too large. Hence in the next numerical computations we will keep $r_c<0.5 R$. 

\section{Radial dependence of deconfinement temperature}

By minimize the total potential $U_{np}+U_p$ at different radial position and temperature, one can find the corresponding $\bar\rho$ as a function of temperature with a certain $\omega$ and radial position. The first temperature at which the $\bar\rho$ deviates from $\pi T$ is the critical temperature, because the $\bar\rho=\pi T$ gives $\langle Tr L\rangle=2cos(\beta \pi T/2)=0$ and thus confinement. As our last work~\ref{Jiang:2023zzu} in order to exhibit the rotation effect more clearly we choose a relatively small radius $R=2(\pi T_0)^{-1}$, $T_0=0.25$GeV and $\omega_{max}=0.9 R^{-1}$. According to Ref.~\cite{Diakonov:2007nv}, $c=2$ and $\Lambda=0.28$GeV in numerical computation. In this work we have not introduced the $\omega$-dependent coupling constant. The main reason is more careful work is needed to get a rotation and position(possible) dependent coupling constant. And this work is in process and more results will be reported in future.

\begin{figure}[ht]
    \centering
    \includegraphics[width=8 cm]{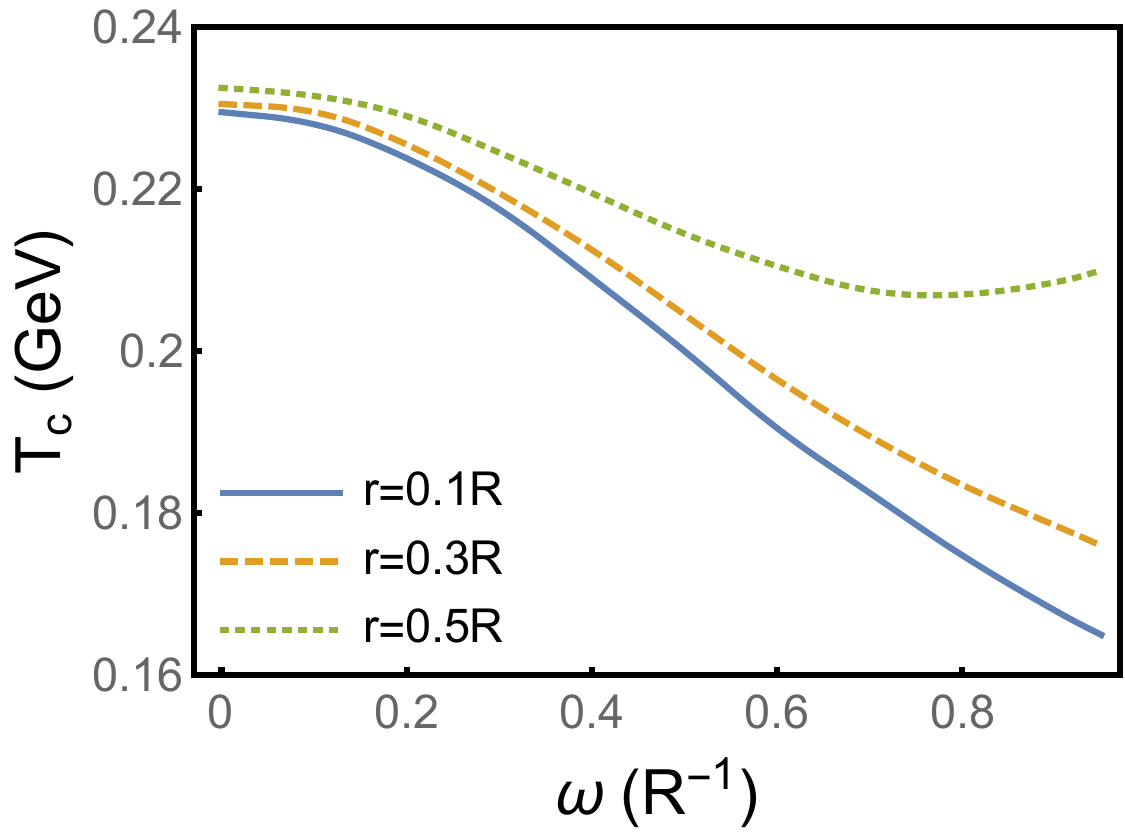}
    \caption{The critical deconfinement temperature as function of a real angular velocity at different radial positions.}
    \label{fig1}
\end{figure}

At different radial position the critical temperature as functions of the real angular velocity is shown in Fig.~\ref{fig1}. Clearly the real rotation will speed up the deconfinement transition no matter where the cell is as the on-axis case. In the $r_c=0.5R$ case, the critical temperature tend to rise around $\omega=0.8 R^{-1}$. It is because the center velocity $u\sim 0.4$ is not quite a small number. Considering the nonlinear formalism of semi-classical potential generates higher power terms of $\omega$, this rising behavior could be quite unphysical. Hence we can focus on the small $\omega$ range. It can be found that the deconfinement temperature becomes higher when the radial distance goes larger. It means the outer part is more difficult to deconfine in the real velocity case. 

\begin{figure}[ht]
    \centering
    \includegraphics[width=8 cm]{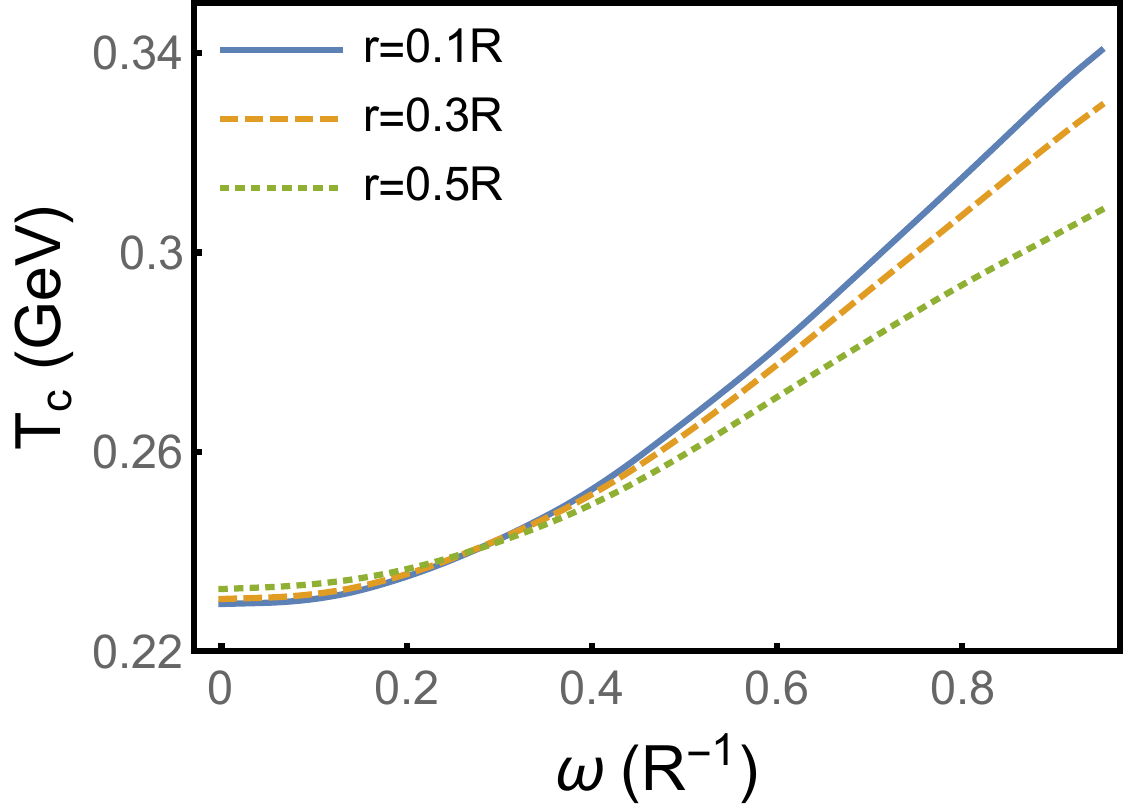}
    \caption{The critical deconfinement temperature as function of a imaginary angular velocity at different radial positions.}
    \label{fig2}
\end{figure}

In order to compare with the lattice result, we also show the imaginary velocity results in Fig.~\ref{fig2}. This is done by the replacement $\omega\rightarrow i\omega$. The main difference in the imaginary case is the radial distance dependent factor $\alpha=1+\omega^2 r_c^2$ always positive. The effective potential can be computed without any technical difficulty at arbitrary $\omega$. It can be found the deconfinement temperature rises when rotation becomes faster. This is qualitatively the same as the on-axis case. When radial distance becomes larger, the critical temperature decreases. This means the outer part is easier to deconfine in the imaginary velocity case. 

According to the recent lattice simulation for a SU(3) gauge field system~\cite{Braguta:2023iyx}, the radial dependence of the deconfinement temperature is the same as results in above for both real and imaginary cases. However the $\omega$ dependence at a certain radial position is contradictory to that in this work. As our last work~\ref{Jiang:2023zzu} indicates the $\omega$-dependent coupling constant will produce the same qualitative behavior as the lattice results. And different from the on-axis case, the critical temperatures at different radial distance are different even $\omega=0$. It is the finite-size effect in the perturbative part which corresponds to $J_n(\xi_n^{(m)}r)$ in Eq.~\ref{pert}.

\section{Conclusions}

In this work we have solved the rotation-modified Yang-Mills equation in the SU(2) case and obtained the rotational corrections to the so-called dyon solutions with arbitrary centers to $\mathcal{O}(\omega^2)$ order. Because of the spherical symmetry breaking in the rotating system, the topological non-trivial solutions can no longer be classified with the homotopy group of the mapping from space-time to group space of SU(2), i.e. $T^3\rightarrow S^3$. Hence there is no strict and spherical symmetric solutions. With these approximated solutions the semi-classical potential has been computed in dilute limit. The radial position dependent deconfinement temperature has been obtained by minimizing the semi-classical potential in both real and imaginary angular velocity cases. Although without considering the $\omega$-dependent coupling constant the critical temperature behaves different from the lattice simulation at each radial position as the rotation goes faster, its radial dependence is qualitatively the same as the lattice. By analyzing the non-perturbative and perturbative potentials separately it is found that the non-perturbative part plays an more important role in the radial dependence of critical temperature because of the inhomogeneous factor $\alpha$. Constrained by the small velocity expansion and the light limit of center velocity in $\alpha$, the computation should stay in the $\omega$ range, or unphysical behavior of critical temperature will emerge. 

In order to compare with the lattice results quantitatively, this work is expected to be improved in two aspects. First is color number of the theory. In SU(3) case there are 3 kinds of dyons, and the asymptotic behavior at spatial infinity are more complicated even in static case. One should consider the SU(3) case seriously to clarify the reason of the contradictory qualitative behaviors of the $\omega$-dependence of critical temperature between dyon ensemble and lattice simulation when no $\omega$-increasing coupling introduced. Second one may go beyond dilute system limit and consider the interaction between dyons and introduce quarks as an even further generalization. By computation the full quantum weight of dyon pairs(caloron) one can obtain the perturbative potential and dyon interaction simultaneously. Such a computation can also improve the result of the perturbative part because the Eq.~\ref{pert} is the same as that with assumption of a constant background color field. Furthermore the lattice simulation~\cite{Braguta:2021ucr} indicates the quarks may provide a contrary contribution to the gluons. This can be checked with the dyon ensemble as well.

\section*{Acknowledgements}
The work of this research is supported by the National Natural Science Foundation of China, Grant Nos. 12375131(YJ). We thanks Xu-Guang Huang very much for his invaluable comments.





\end{document}